# INSIGHT ABOUT DETECTION, PREDICTION AND WEATHER IMPACT OF CORONAVIRUS (COVID-19) USING NEURAL NETWORK

A K M Bahalul Haque, Tahmid Hasan Pranto,
Abdulla All Noman and Atik Mahmood

Department of Electrical and Computer Engineering,
North South University, Dhaka-1229, Bangladesh

## ABSTRACT

*The world is facing a tough situation due to the catastrophic pandemic caused by novel coronavirus (COVID-19). The number people affected by this virus are increasing exponentially day by day and the number has already crossed 6.4 million. As no vaccine has been discovered yet, the early detection of patients and isolation is the only and most effective way to reduce the spread of the virus. Detecting infected persons from chest X-Ray by using Deep Neural Networks, can be applied as a time and labor-saving solution. In this study, we tried to detect Covid-19 by classification of Covid-19, pneumonia and normal chest X-Rays. We used five different Convolutional Pre-Trained Neural Network models (VGG16, VGG19, Xception, InceptionV3 and Resnet50) and compared their performance. VGG16 and VGG19 shows precise performance in classification. Both models can classify between three kinds of X-Rays with an accuracy over 92%. Another part of our study was to find the impact of weather factors (temperature, humidity, sun hour and wind speed) on this pandemic using Decision Tree Regressor. We found that temperature, humidity and sun-hour jointly hold 85.88% impact on escalation of Covid-19 and 91.89% impact on death due to Covid-19 where humidity has 8.09% impact on death. We also tried to predict the death of an individual based on age, gender, country, and location due to COVID-19 using the LogisticRegression, which can predict death of an individual with a model accuracy of 94.40%.*

## KEYWORDS

## 1. INTRODUCTION

A pneumonia of unknown type was detected in Wuhan, China on 31 December 2019. [1] On 11 February 2020, WHO announced a name for the new coronavirus disease: COVID-19. [1] Covid-19, the super spreader virus that is causing the 2019-2020 coronavirus global pandemic, has already become a great threat to the human race. As of 05 June 2020, it has already taken 393254 lives and over 6.7 million cases have been confirmed from 215 countries. [2] Coronavirus comes from a family named coronaviridae which includes more than 40 similar kind of viruses like SARS, MARS and Covid-19. [3] Coronaviruses are there from 1960's. [4] Before coronavirus, there was a SARS-CoV outbreak in 2003 at Mainland China and Hong Kong. And in 2012, MARS-CoV out broke in Saudi Arabia, UAE, Korea and other countries. [4] But unlike SARS or MARS, the human to human transmission and human death rate due to covid-19 is incomparable. The coronavirus enters a human body from droplets, cough or direct contact, takes down our immunity system and does momentous damage to our lungs. [5]

As there is no official medication of this virus, it has created a public panic resulting in economic and social anarchy. This virus has a very high substitution rate, they mutate rapidly and also very susceptible to recombination. [6] Another characteristic that makes this virus even more dangerous is its mutation. [7] The virus has already mutated several times and no one can tell how many times it will mutate and get stronger. To fight this global pandemic, detecting the patient and isolation of that patient is known to be the most effective primary level of treatment. As many countries around the globe are lacking Covid-19 detection kit, using technology for detection can be a solution to be considered. Deep neural networks and Machine Learning is a reliable technology that can detect Covid-19 from chest x-ray images within seconds. The standard covid-19 detection test, PCR is short in supply across the world which also requires high sensitivity and time for detection. But as the pandemic is growing exponentially, we need faster ways to detect the patients and for that, technology like machine learning, deep learning can be used as an alternative to clinical tests. The use of machine learning and neural networks has already brought revolutionary changes in medical and health science in the last decade. The use of machine learning algorithms will save the valuable time of medical professionals.

Keeping these things in mind, we decided to use machine learning models to analyze the situation as per weather condition and detect covid-19 from x-ray images and also on the basis of age, gender, city and country our model will predict the person's chances to survive this pandemic. We took some pre-trained machine learning models that are well known for their image classification capability as the input layer of our machine learning architecture and then as the processing layer we took some dense layers and finally the output layer successfully classifies the result between normal x-ray, pneumonia and covid-19, that is what kind of infection that the person is affected with. And for prediction and analysis we used machine learning regression models.

In this paper first we have a discussion about the most recent works in regarding this issue in section 2. Then in section 3 we will explain how we collected our necessary data, processed them and the methodologies that we used. Section 3 will also include a brief discussion about the architecture of our machine learning models that we have used for both detection and analysis. Then we will analyze our results and findings in section 4. The later part of this paper terminates the topic while discussing the shortcomings that we have faced during this research.

## 2. RELATED WORKS

Machine learning and artificial intelligence has already become popular for its prediction and detection capabilities. The immense development of AI, machine learning and deep learning is showing us a new future where these three technologies will be used in almost every sector.[8] A greater part of medical science has already adopted machine learning and artificial intelligence for analyzing the patient's data, detecting and predicting disease and that too with a great precision and reliability. Adoption of machine learning is saving the valuable time of the technicians as well as other medical workers. [9] From prevention of healthcare problems to detection of post-operative infections, machine learning can be used in almost all aspects. [10, 11]

WHO has suggested to test the person who has a travel history recently or has been in close contact with any person who has been detected positive with Covid-19. For detecting Covid-19 with the help of machine learning, using X-Ray images is the most preferable way. Very deep neural network along with pre-trained machine learning models can be a very reliable and powerful classification tool for classifying affected X-Rays. [12] By using some X-Ray images labeled with assurance from medical technicians, machine learning models can be trained and used later on for automatic detection while saving valuable time. The medical process PCR is not 100% accurate, the sensitivity level of PCR is 60-70% while machine learning models can perform similar to that, in some cases, even much better than PCR. [13] X-Ray and CT-images combined with clinical image features provides valuable information that can detect Covid-19 in the early developing stage where PCR will give a false negative result. [13]

Many previous researches have established the impact of weather with the escalation of viral disease. However, no absolute relation has been established between Covid-19 and weather elements. But temperature and humidity have shown relation with Covid-19. [14]

## 3. MATERIALS AND METHODS

The materials we needed includes datasets (x-ray dataset, weather dataset and a complete Covid-19 dataset) and the tools we used includes some python libraries (numpy, pandas, scikit learn), for the machine learning backend, we used tensorflow and for the pre-trained models, we used keras as keras has some built in pre-trained deep neural model for classification.

### 3.1 Dataset

We used a total of five datasets. Table 1 contains the dataset handles of the datasets that we used.

Table 1: Dataset handles.

| Task | Used Datasets |
| --- | --- |
| 1. Weather Impact analysis | 1. Covid-19 Complete Dataset (Updated Every 24 Hours) [15]<br><br>2. Covid-19 Global Weather Data [16] |
| 2. Covid-19 Detection | 1. Kaggle Chest X-Ray Images. (Pneumonia) [17]<br><br>2. Covid Chest X-Ray Dataset. [18] |
| 3. Predicting Death | 1. Novel Coronavirus 2019 Dataset. [19] |

Figure 1 shows some representative images that we used for covid-19 detection, table 2 shows overview of data we used for weather factor analysis and table 3 shows data overview that was used for death prediction.

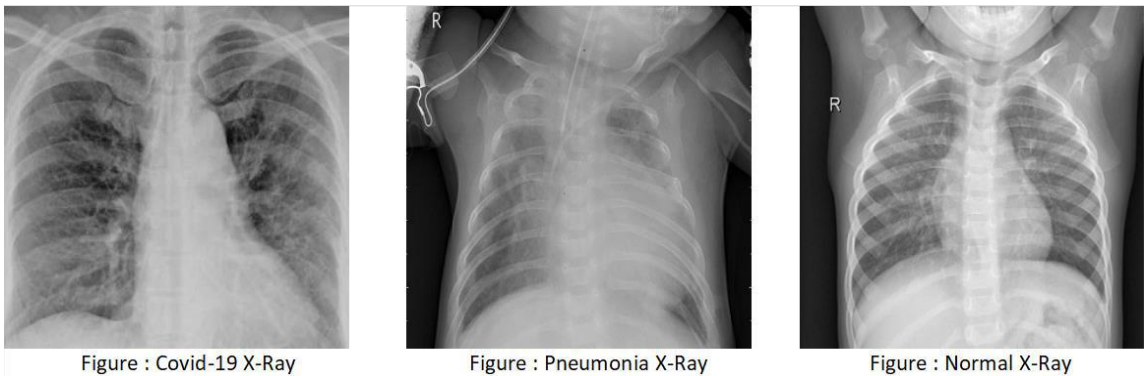

Figure : Covid-19 X-Ray            Figure : Pneumonia X-Ray            Figure : Normal X-Ray

Figure 1: Representative image from chest X-Ray dataset.

Table 2: Weather factor impact analysis data.

| confirmed | death | humidity | tempC | sunHour | windspeedKmph |
|----------:|------:|---------:|------:|--------:|--------------:|
| 43 | 0 | 63.0 | 18.0 | 6.1 | 8.0 |
| 70 | 0 | 75.0 | 25.0 | 8.8 | 6.0 |
| 94 | 0 | 81.0 | 30.0 | 8.9 | 12.0 |
| 2 | 0 | 85.0 | 21.0 | 7.4 | 9.0 |
| 0 | 0 | 52.0 | 18.0 | 11.6 | 8.0 |

Table 3: Death prediction data.

| location | country | gender | age | death |
|---------:|--------:|-------:|----:|------:|
| South Korea | South Korea | male | 24 | 0 |
| South Korea | South Korea | male | 24 | 0 |
| Wan Chai | Hong Kong | male | 24 | 0 |
| Hong Kong | Hong Kong | male | 24 | 0 |
| Kyoto | Japan | female | 25 | 0 |

## 3.2 Data Process

The x-ray images were scattered among folders. We combined them as per as the need and inserted them in two folders for train and test data. All the images were resized to 150x150 during loading dataset into the codebase. And we used another csv files which contained data of about 1086 death cases with respect to the persons gender, country, age and location. To get the best possible prediction we dropped all the incomplete rows. After cleaning the dataset, we were left with data of 831 people. And for the weather factors analysis we combined two datasets and made a data-frame which contains date wise data like humidity, temperature, sun hour and wind speed. There was a total of 16677 rows of data grouped by date, country and city. The whole data process and training for weather impact analysis and death prediction follows the high-level architecture shown in figure 2.

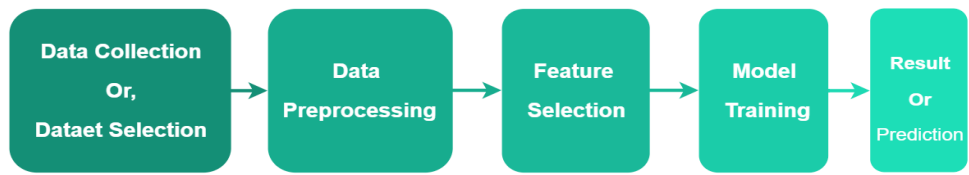

Figure 2: High level data process architecture.

## 3.3 Tools

TensorFlow [20] was used as the backend of all the machine learning related works related to our research. We used python for coding and python distribution of some tools like Scikitlearn [21], Pandas [22], NumPy [23] or matplotlib [24] for different implementations. Anaconda with jupyter notebook was used as the environment. Keras pre trained models were used. All the references of pre trained models are stated on Keras documentation [25]. All the computation was done using a laptop that is Asus ZenBook 15 (16GB Ram/ 1Tb PCIe SSD, 8GB graphics) equipped with a GTX 1050(Max-Q) 2Gb external graphics. Windows 10 home OS was used during the research.

## 3.4  Methods

### 3.4.1  Weather impact analysis

We used DecisionTreeRegressor to analyze the weather factor impacts on Covid-19 escalation. Decision tree builds regression models in the form of a tree structure. It breaks down a dataset into smaller and smaller subsets while at the same time an associated decision tree is incrementally developed. The final result is a tree with decision nodes and leaf nodes. [26] Before processing the data, we observed that necessary data was in two separate datasets. From the Covid-19 Complete Dataset [15] we took country, state, death, recovered, confirmed cases. Then we added active cases column in that where,

Active cases = confirmed – (death + recovered)

Then from the Covid-19 Global Weather Dataset [16], we took country, state, date, temp, windspeed, sun hour, humidity columns. Then these two datasets were merged on basis of country, state and date. We dropped all the data that wasn't important for this research. We took temperature, humidity, sun hour and wind speed as our feature data and death and confirmed cases as our labeled data. 80% data was used for training and the other 20% data was used as test data. The impact of weather factors both on death due to covid-19 and the escalation of covid-19 is discussed in the result and analysis section (Section 5).

### 3.4.2  Detection using X-Ray

For detection, we used convolutional neural network. As the input layer we used keras pre-trained machine learning models like VGG16, VGG19, InceptionV3, Xception and Resnet50. [25] And in all the cases the inner layers of our model started with a flatten layer followed by some full connected dense layers. The dense layers successively contain 512, 256, 128, 64 numbers of neurons. These four layers uses rectified linear unit as the activation function. The outermost layer contains three neurons as we have 3 classes to classify. The outer layer uses SoftMax activation function. The structure is shown in figure 3 below. Two hundred images were used as training image and the model was tested on 20 images. The train and validation results are discussed in section 4.

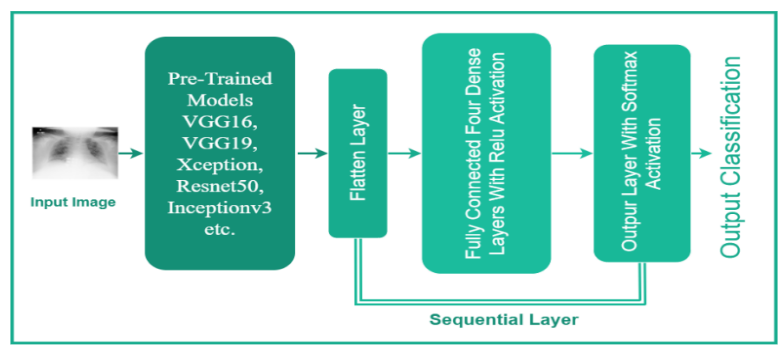

Figure 3: Structure of X-Ray classification CNN.

Section 4 also compares and analyzes the outputs when we vary the pre-trained models in the input layer. All the images were trained for 20 epochs and batch size of 10. Images were augmented during training. All of them was trained with loss function named "categorical-crossentropy" and the optimizer was "Adam" with a 0.0005 learning rate for all of them.

### 3.4.3 Death Prediction

For death prediction we used Logistic Regression machine learning model. A Logistic Regression model can handle prediction over multiclass level. [27] In the multiclass case, the training algorithm uses the one-vs-rest (OvR) scheme. The data used for prediction contained object type data, which was uncompilable for the model. We used Scikit-Learn Lebel Encoder to decode the features so that we can compile the model. Gender, age, location and country was our feature and death were the label. Given a person's gender, age, location and country, this model can determine the persons chance of surviving during this pandemic with a model accuracy score of 94.40%. With more data, this model is expected to perform better in real life scenario.

## 4. RESULT & ANALYSIS

### 4.1 Weather Impact:

The two datasets we used to analyze weather impact contains many insights. We tried to find the death per one thousand ranking. The ranking is shown in figure below marked as figure 4. As we can see from the horizontal bar diagram, the death rate per thousand in Belgium is the highest according to the dataset that we used. But, is there any relation between the death rate and weather? Or is there any relation with escalation of Covid-19 with weather? We tried to find out the answer to these questions.

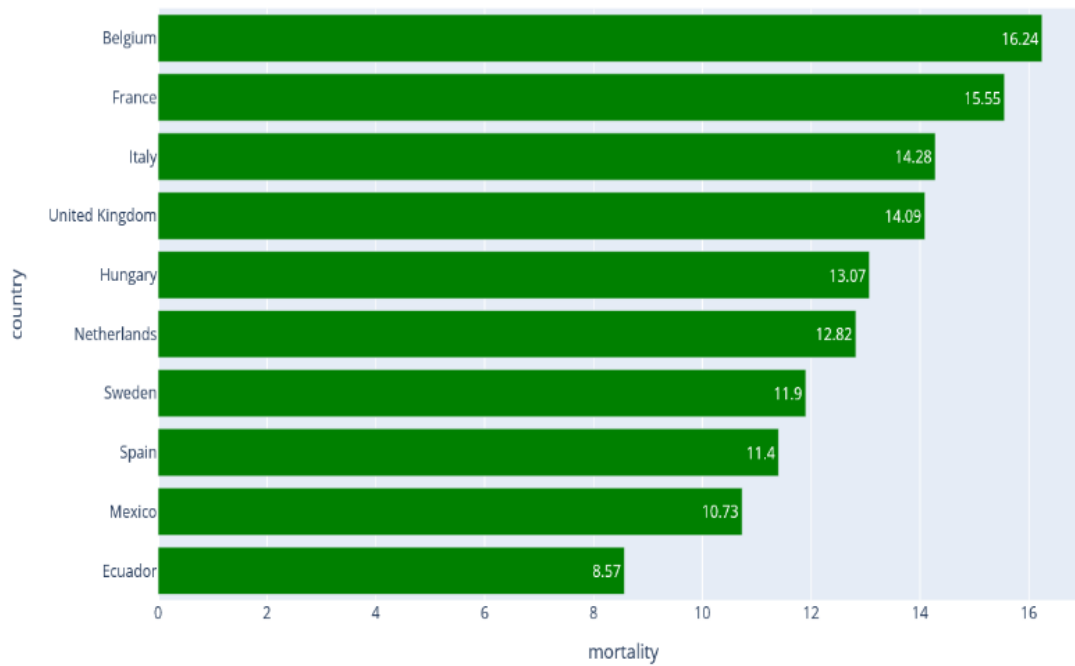

Figure 4: Mortality rate per thousand cases.

The datasets we merged together to analyze the weather factor impact, contains information of 188 countries and this data is was grouper together by date, country and state. There is weather data of 131 days, using these data we analyzed the percentage impacts of weather factors on Covid-19. We trained two DecisionTreeRegressor models to find out the associated percentage impact of weather factors with death and spread of Covid-19. The first DecisionTreeRegressor determines the impact percentages of weather factors in the escalation of Covid-19 which is shown in figure 5. And the second DecisionTreeRegressor determines the percentage impact of weather factors on death due to Covid-19 which is shown in figure 6.

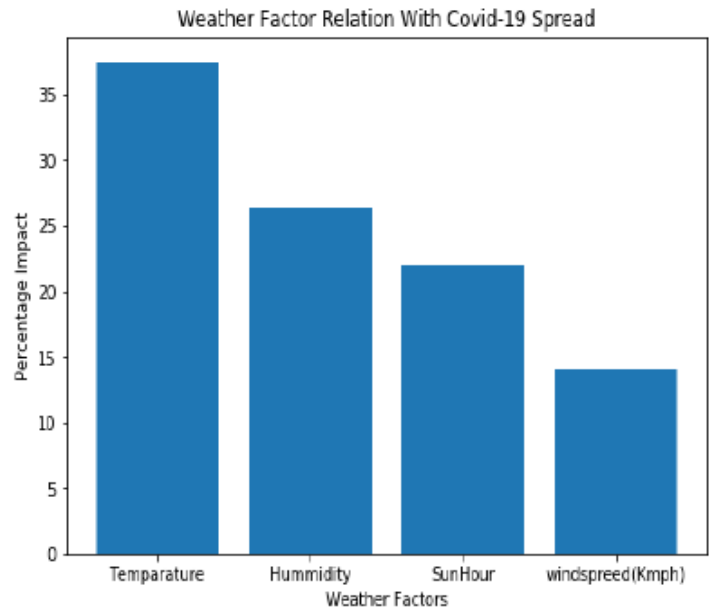

Figure 5: Weather factor impact on escalation

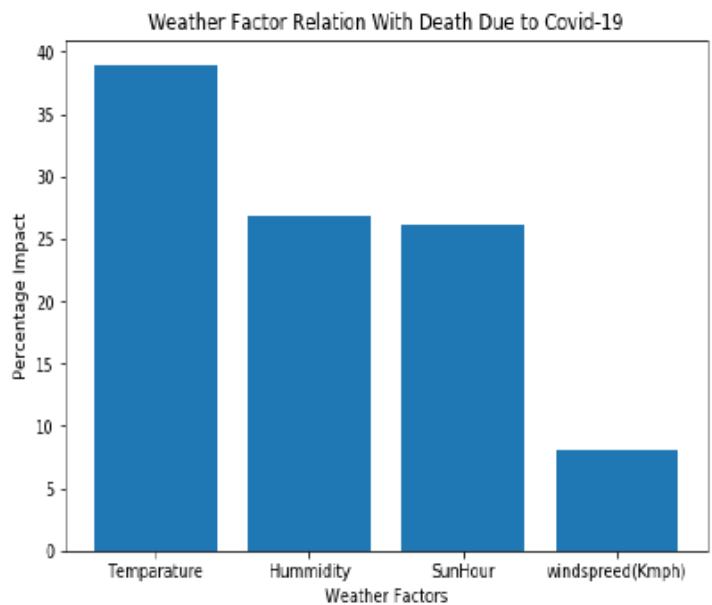

Figure 6: Weather factor relation with death

From figure 5 we can see that the impact of temperature in escalation of Covid-19 is 37.51%, impact of humidity is 26.32%, impact of sun hour is 22.05% and impact of wind speed is 14.10%. So, it can be stated that, temperature and humidity have a direct and strong relation with escalation of Covid-19. A total of 63.83% of impact is very significant where it concerns exponential growth of a virus worldwide, taking so many lives.

The other Decision Tree Regressor model determines the impact of weather factors on death due to Covid-19 which is shown in figure 6. As we can see from the graph temperature has 38.99% impact, humidity has 26.80% impact, sun hour has 26.10% impact and wind speed has 8.09% impact on deaths due to covid-19. From figure 5 we have already seen that temperature and humidity have 63.83% impact on Covid-19 escalation and figure 6 shows that temperature and humidity have 65.79% impact on death. 65.79% impact is very significant if it concerns death which is the case we tried to find.

Combining all the findings we got from these graphs, in both escalation and death, temperature holds the highest impact that is 37.51% in escalation and 38.99% in death. So, the relation of temperature with Covid-19 is clearly seen and understood. This can be clearly said that temperature have a definite and significant relation with deaths due to Covid-19 and spread of Covid-19.

## 4.2 Death Prediction

Death was predicted on basis of age, gender, country and state. Logistic Regressor first multiplies a randomized weight matrix with the features to create a linear model.

**Linear Model:** $a = w_0 + w_1x_1 + w_1x_2 + \cdots + w_nx_n$ (equation-1)

Then cost functions adjusts these costs to get the best possible result. Finally, sigmoid like function is applied as we are predicting if a person will survive or not. This function keeps the values between 1 and 0.

**Logistic Model:** $p = \frac{1}{1+e^{-a}}$ (a is from equation-1)

To determine the result, the co-efficient values holds a great impact. The coefficient values are shown in figure 7.

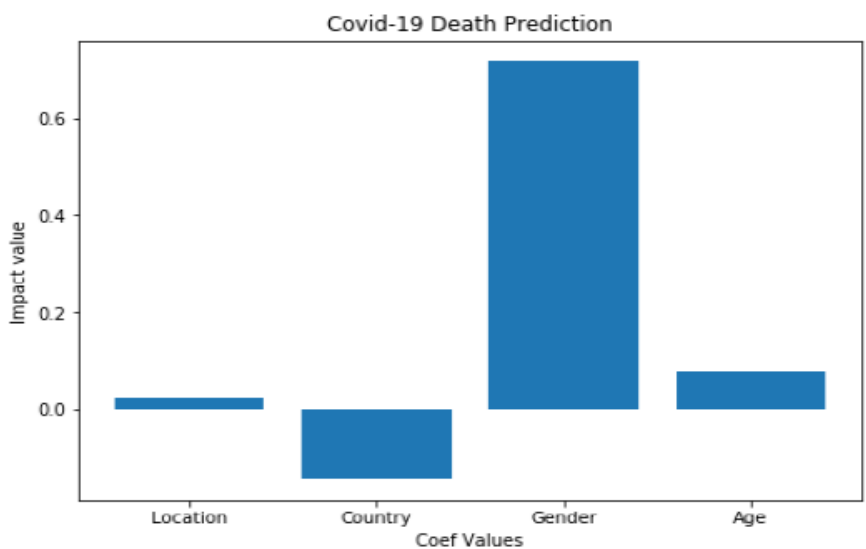

Figure 7: Coefficient values and impact.

From the figure 7 it is can be said that, death is mostly related to the persons gender and age. On the prediction, among the 4 features, gender has the highest effect on death. As study already shows men are more likely to die due to Covid-19. [28]

Figure 8 shows a graphical representation of prediction by the model. Tough it can be clearly seen that there is a false prediction in the graph, but, the performance of the model can easily be understood by the confusion matrix shown in figure 9.

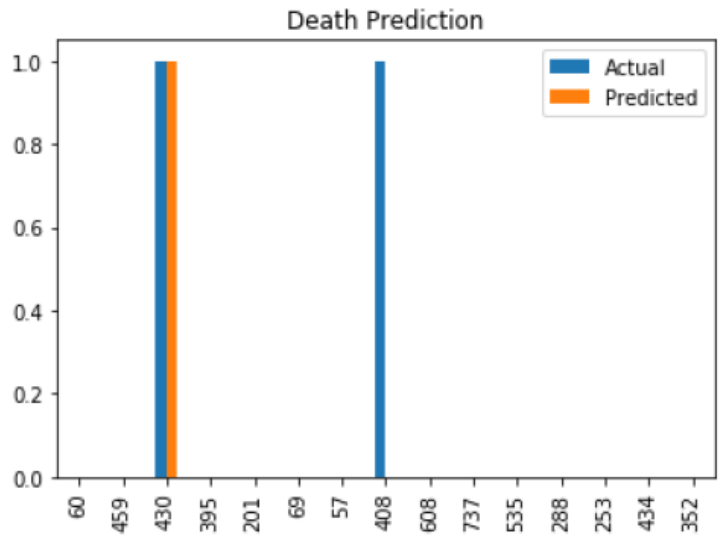

Figure 8: Death prediction

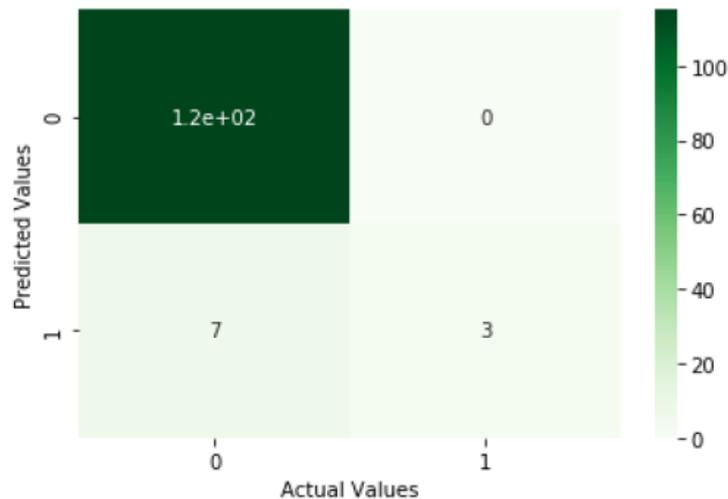

Figure 9: Confusion Matrix

There was a total of 125 test data. The confusion metrics gives us the insight about how many correct and false prediction the model has made. From the heatmap in figure 9, we can analyze the prediction. There is 4 possible prediction that the model can make. TP (True Positive), TN (True Negative), FP (False Positive) and FN (False Negative). Among 125 data, our model has predicted 115 TP values and 3 TN which are correct predictions. The model only made 7 FN values which are wrong predictions. The effectivity of this machine learning based prediction system can be more clearly understood by sensitivity, specificity, precision, F1-score.

$$\text{Sensitivity} = \frac{TP}{TP+FN} = \frac{115}{115+7} = 94\%$$

$$\text{Specificity} = \frac{TN}{TN+FP} = \frac{3}{3+0} = 100\%$$

$$\text{Precision} = \frac{TP}{TP+FP} = \frac{115}{115+0} = 100\%$$

$$\text{f1-score} = \frac{2*Precision*Sensitivity}{Precision+Sensitivity}$$

$$= \frac{2*100*94}{100+94} = 96.90\%$$

From the scores along with the accuracy of 94.04% proves that the model is predicting with a very adequate sensitivity, specificity and precision.

Although the data wasn't sufficient enough, the model showed promising performance in prediction.

## 4.3 Covid-19 Detection from X-Ray

X-ray images showed rapid progression of pneumonia and some differences between the left and right lung. [29] So deep learning can learn from that differences in X-Ray images. By varying the first input layer of our system we tested the performance of various machine learning pre-trained models which are known for their image classification capability. But, all of them didn't work well with our dataset. Figure 10 shows the training accuracy of each model after 20 epochs. And figure 11 shows the validation accuracy of models after 20 epochs.

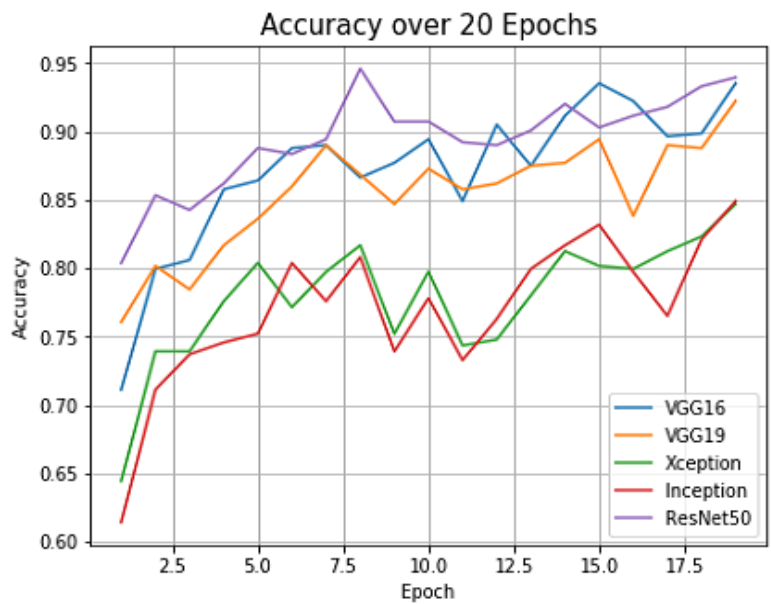

Figure 10: Training accuracy over 20 epochs

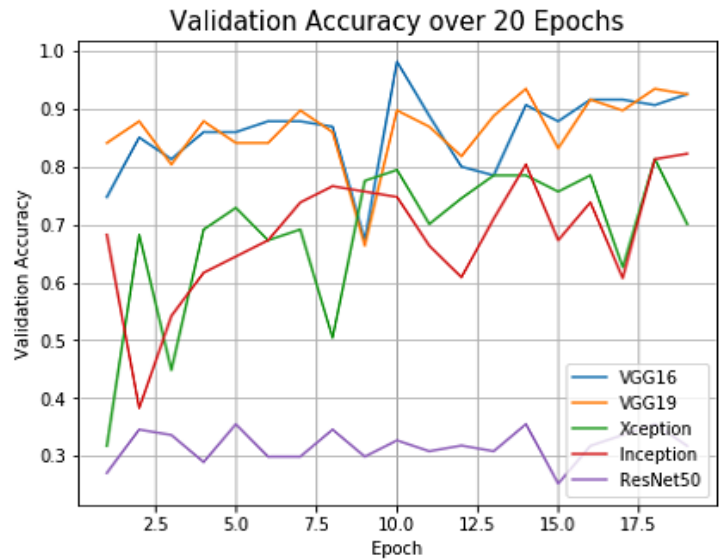

Figure 11: Validation Accuracy over 20 epochs

From the accuracy graphs we can see the VGG16, VGG19 and Xception has an increasing accuracy rate. Increase in the accuracy rate proves that the model is learning. We can also see the similar characteristics in the loss functions. VGG16 and VGG19 has the lowest loss. Figure 12 shows training loss over 20 epochs and figure 13 shows validation loss over 20 epochs.

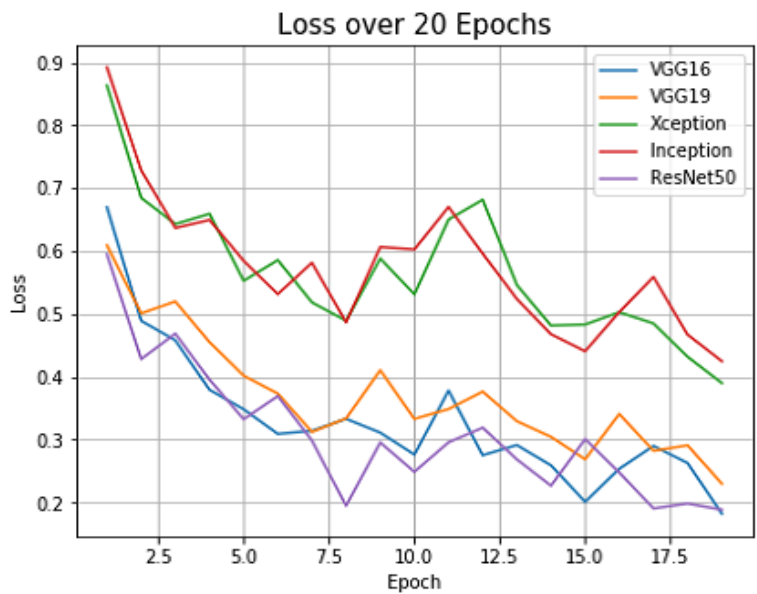

Figure 12: Training loss over 20 epochs

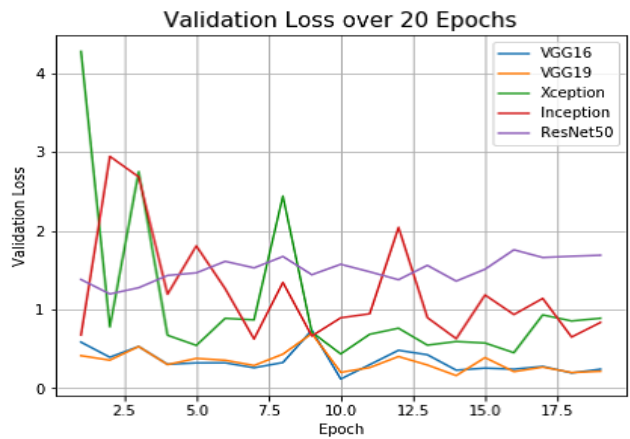

Figure 13: Validation Loss over 20 Epochs

The train, test accuracy and loss scores are given below in table 4.

| Accuracy and loss after 20 epochs` | | | | |
|---|---|---|---|---|
| **Name** | Accuracy | Validation Accuracy | Loss | Validation Loss |
| **VGG16** | 0.9362 | 0.925 | 0.1805 | 0.2417 |
| **VGG19** | 0.9234 | 0.925 | 0.2275 | 0.2148 |
| **Xception** | 0.8489 | 0.700 | 0.3896 | 0.8878 |
| **InceptionV3** | 0.8448 | 0.822 | 0.4288 | 0.8346 |
| **ResNet50** | 0.9404 | 0.317 | 0.1865 | 1.6878 |

Table 4: Accuracy and loss values using different pre-trained keras models.

From the table-4 we can see that VGG19 and VGG16 performing very well with unknown data means the model has learnt from the given data. But despite having a higher accuracy of about 94%, ResNet50 haven't learnt anything from the given data and it can be seen from the validation accuracy of 31% which is very low. And the same scenario is seen in the loss function as well. VGG16 and VGG19 has the lowest loss and ResNet50 has the highest loss. The other model's despite of having decent accuracy, will not perform well for either having a high loss or having a huge difference between training and validation accuracy.

## 5. FUTURE WORK & CONCLUSION

Considering the current situation due to coronavirus outbreak, we tried to analyze the situation, detect the disease and predicted the persons chances of survival. This is by no means an industry ready production but the models we built showed promising result as per detection and prediction scores. For detection, VGG16 and VGG19 performed significantly better than other models. We analyzed the weather factor and found that temperature has a direct and significant relation with death caused by covid-19 and in the escalation of it too. For prediction and analysis, we used selected regressor models as per the nature of data and type of task we wanted to conduct. Which also showed promisingly accurate result also. As the pandemic is quite new, data is not enough. With more data, the models are anticipated to work better.

These models can be integrated in hospital websites and can be used by doctors to detect the patients much faster.

## 6. DECLARATIONS

### 6.1 Study Limitations

More X-Ray images labeled by expert medical technicians and doctors will increase the reliability of the neural network model. More clear and high definition X-Ray images would have made difference in the accuracy of the covid-19 detection neural model. The dataset for death prediction wasn't sufficient enough to make an absolute prediction though the model performed promisingly well.

### 6.2 Future Scope of Improvements

The model to detect covid-19 basically classifies the infected X-Rays and other X-Rays. This very similar model can be used for other classification or identification-based studies as well. Any X-Ray based detection can be done using the same architecture that we have used. We only need labeled data to train the models so that it can learn from the data. Deployment of deep learning models in real hospital scenario will increase the scope of research on this area.

### 6.3 Acknowledgements

We would like to express our very great appreciation to our supervisor Mr. A. K. M. Bahalul Haque for his consistent support throughout the planning and development of this research. Advices provided by him has been a great help to us.

## 7. COMPARISON BETWEEN THIS WORK AND RELATED WORKS

| Paper Name | Findings/Outcome | Relevance with Our Work |
|---|---|---|
| 1. Artificial Intelligence and Its Impact on The Fourth Industrial Revolution: A Review | 1. This paper shows how development trends in AI, Machine Learning and Deep Neural Network is going to change the future. | 1. We are using machine learning and deep neural network for detection and prediction. |
| 2. Machine learning for medical diagnosis: history, state of the art and perspective. | 2. This paper shows the overview of machine learning based intelligent data process in medical sector. | 2. Our work also processes data with machine learning to predict death and also to analyze weather impact related to covid-19 pandemic. |
| 3. Prevention of heart problems using artificial intelligence. | 3. Shows how data analysis can predict heart attack. | 3. Prediction is one of the main features of our system. |
| 4. An intelligent model for detection of post-operative infections. | 4. This research uses machine learning and processes medical data to extract most significant feature for detecting post-operative infection. | 4. Our work uses feature importance to analyze the weather impact. |
| 5. COVID-19 Detection using Artificial Intelligence. | 5. This research classifies covid-19 X-Rays between two classes. | 5. Our work classifies X-Rays between three classes using pre-trained deep learning models. |

| 6. Automated detection of COVID-19 cases using deep neural networks with X-ray images. | 6. Classifies Coid-19 X-Rays with 87.02% accuracy for multi class. | 6. Our model Classifies with 93.6% accuracy in multi class level. |
|---|---|---|
| 7. The role of environmental factors on transmission rates of the COVID-19 outbreak: An initial assessment in two spatial scales. | 7. This research shows rapid growth of coronavirus over a range of temperature. | 7. We have shown what weather factors impacts the growth of coronavirus and also the impacts the death rate. |